\documentstyle[12pt,psfig]{article}

\def\thefootnote{\fnsymbol{footnote}}

\begin{document}

\newcommand{\gsim}{ \mathop{}_{\textstyle \sim}^{\textstyle >} }
\newcommand{\lsim}{ \mathop{}_{\textstyle \sim}^{\textstyle <} }
\newcommand{\vev}[1]{ \left\langle {#1} \right\rangle }
\newcommand{\lsp}{ \left ( }
\newcommand{\rsp}{ \right ) }
\newcommand{\lmp}{ \left \{ }
\newcommand{\rmp}{ \right \} }
\newcommand{\llp}{ \left [ }
\newcommand{\rlp}{ \right ] }
\newcommand{\labs}{ \left | }
\newcommand{\rabs}{ \right | }
\newcommand{\K}{ {\rm K} }
\newcommand{\EV}{ {\rm eV} }
\newcommand{\KEV}{ {\rm keV} }
\newcommand{\MEV}{ {\rm MeV} }
\newcommand{\GEV}{ {\rm GeV} }
\newcommand{\TEV}{ {\rm TeV} }
\newcommand{\citen}[1]{ {\cite{#1}} }

\newcommand{\Journal}[4]{{\sl #1} {\bf #2} {(#3)} {#4}}

\newcommand{\APJ}{Ap. J.}
\newcommand{\APP}{Astroparticle Phys.}
\newcommand{\CJP}{Can. J. Phys.}
\newcommand{\NC}{Nuovo Cimento}
\newcommand{\NP}{Nucl. Phys.}
\newcommand{\PL}{Phys. Lett.}
\newcommand{\PR}{Phys. Rev.}
\newcommand{\PRep}{Phys. Rep.}
\newcommand{\PRL}{Phys. Rev. Lett.}
\newcommand{\PTP}{Prog. Theor. Phys.}
\newcommand{\SJNP}{Sov. J. Nucl. Phys.}
\newcommand{\ZP}{Z. Phys.}

\begin{titlepage}
\begin{center}

\hfill    LBL-37911\\

\vskip .5in

{\Large \bf
On the Solution to the Polonyi Problem with No-Scale Type
Supergravity\footnote
{Talk given at the Yukawa International Seminar '95 in Kyoto, 21 -- 26
August, 1995. This talk is based on the works in collaborations with
M.~Kawasaki, M.~Yamaguchi and T.~Yanagida.}
\footnote{This work was supported by the Director, Office of Energy
Research, Office of High Energy and Nuclear Physics, Division of High
Energy Physics of the U.S. Department of Energy under Contract
DE-AC03-76SF00098.}
}

\vskip .5in

{T.~Moroi}

\vskip .5in

{\it
Theoretical Physics Group, Lawrence Berkeley National Laboratory\\
University of California, Berkeley, CA 94720, U.S.A.\\
and\\
Theory Group, KEK, Ibaraki 305, Japan
}

\end{center}

\vskip .5in

\begin{abstract}

We study the solution to the Polonyi problem in the framework of
no-scale type supergravity. In such a model, Polonyi field can weigh
as $O(10{\rm TeV})$ and decay just before the big-bang
nucleosynthesis. It is shown that in spite of a large entropy
production by the decay of the Polonyi field, one can naturally
explain the present value of the baryon-to-entropy ratio, $n_B/s \sim
(10^{-10}-10^{-11})$ if the Affleck-Dine mechanism for baryogenesis
works. It is pointed out, however, that there is another cosmological
problem related to the abundance of the lightest superparticles
produced by the decay of the Polonyi field.

\end{abstract}
\end{titlepage}
\renewcommand{\thepage}{\roman{page}}
\setcounter{page}{2}
\mbox{ }

\vskip 1in

\begin{center}
{\bf Disclaimer}
\end{center}

\vskip .2in

\begin{scriptsize}
\begin{quotation}
This document was prepared as an account of work sponsored by the United
States Government. While this document is believed to contain correct
information, neither the United States Government nor any agency
thereof, nor The Regents of the University of California, nor any of their
employees, makes any warranty, express or implied, or assumes any legal
liability or responsibility for the accuracy, completeness, or usefulness
of any information, apparatus, product, or process disclosed, or represents
that its use would not infringe privately owned rights.  Reference herein
to any specific commercial products process, or service by its trade name,
trademark, manufacturer, or otherwise, does not necessarily constitute or
imply its endorsement, recommendation, or favoring by the United States
Government or any agency thereof, or The Regents of the University of
California.  The views and opinions of authors expressed herein do not
necessarily state or reflect those of the United States Government or any
agency thereof, or The Regents of the University of California.
\end{quotation}
\end{scriptsize}

\vskip 2in

\begin{center}
\begin{small}
{\it Lawrence Berkeley Laboratory is an equal opportunity employer.}
\end{small}
\end{center}

\newpage
\renewcommand{\thepage}{\arabic{page}}
\setcounter{page}{1}

\renewcommand{\thefootnote}{\arabic{footnote}}
\setcounter{footnote}{0}

\section{Introduction}
\label{sec:intro}

$N=1$ supergravity~\cite{sugra} is not only regarded as an effective
field theory of superstring below the Planck scale, but also provides
a natural framework for the origin of the soft supersymmetry
(SUSY)-breaking terms. Most of the supergravity models, however,
contain a light massive boson $\phi$ (Polonyi field) with the mass
$m_\phi$ of order the gravitino mass
$m_{3/2}$~\cite{PLB131-59,PRD49-779,PLB318-447}, which is responsible
for the spontaneous SUSY breaking. The Polonyi field $\phi$ couples
only gravitationally to the light particles and hence the lifetime of
$\phi$ is very large as
\begin{equation}
    \tau_\phi \simeq
    \Gamma_\phi^{-1}
    \sim \left( N\frac{m_\phi^3}{M_P^2}\right)^{-1},
    \label{decay_rate}
\end{equation}
where $\Gamma_\phi$ is the decay rate of the Polonyi field,
$M_P=\sqrt{8\pi}M \simeq 1.2\times 10^{19}{\rm GeV}$ the Planck mass,
and $N$ the number of the decay modes. (In the following calculations,
we take $N=100$.)  Then, the Polonyi field is expected to decay when
the temperature of the universe becomes very low. The reheating
temperature $T_R$ due to the decay of the Polonyi field is given by
\begin{equation}
T_R \sim 1 {\rm MeV} \left(\frac{m_{\phi}}{10{\rm TeV}}\right)^{3/2}.
\label{rtemp}
\end{equation}

This fact leads to a serious cosmological difficulty (so-called
Polonyi problem)~\cite{PLB131-59,PRD49-779,PLB318-447}.  Under quite
general assumptions, the Polonyi field $\phi$ takes an amplitude of
order $M$ at the end of inflation, and subsequently it starts
oscillation and dominates the energy density of the universe until it
decays. If the decay of the Polonyi field occurs after or during the
big-bang nucleosynthesis, it most likely destroys one of the
successful scenarios in the big-bang cosmology, that is the
nucleosynthesis.  Furthermore, the decay of the Polonyi field releases
a tremendous amount of entropy and dilutes primordial baryon asymmetry
much below what is observed today. Especially, the important point is
that we cannot solve this problem even if we assume an inflation,
which is the crucial difference between the Polonyi problem and
another serious cosmological difficulty in $N=1$ supergravity, {\it
i.e.} the gravitino
problem~\cite{PRL48-1303,PLB145-181,PLB158-463,PLB189-23,PLB303-289,PTP93-879}.

It has been pointed out~\cite{PLB174-176} that the first problem can
be solved by raising the Polonyi mass $m_\phi$ (or equivalently the
gravitino mass $m_{3/2}$) up to $O(10{\rm TeV})$ so that the reheating
temperature $T_R$ by the decay of the Polonyi field becomes larger
than $O(1{\rm MeV})$.  Then, the nucleosynthesis may re-start after
the decay of the Polonyi field.  This solution favors strongly
``no-scale type'' supergravity~\cite{PRD50-2356},\footnote
{In the original no-scale supergravity
model~\cite{PLB143-410,PLB147-99}, Polonyi field acquires a mass of
the order of $m_{3/2}^2/M$~\cite{NPB241-406} which is much smaller
than the gravitino mass $m_{3/2}$. However, in the ``no-scale type''
supergravity model studied in Ref.~\citen{PRD50-2356}, the mass of the
Polonyi field is at the order of the gravitino mass.}
since the gravitino mass can be taken $O(10 {\rm TeV})$ without
diminishing the original motivation of SUSY as a solution to the
hierarchy problem~\cite{Maiani,Veltman}.  Namely, we can raise the
gravitino mass while keeping all masses of SUSY particles in
observable sector to be $O(100\GEV)$.

Here, we stress that the second problem can be also solved if the
Affleck-Dine mechanism~\cite{NPB249-361} for baryogenesis works in the
early universe~\cite{PLB342-105}. However, we point out another
cosmological problem that the lightest superparticles (LSPs) produced
via the Polonyi decay are extremely
abundant~\cite{PLB342-105,lbl37715}. As a result, their energy
density, if stable, overcloses the universe unless the reheating
temperature due to the Polonyi decay is sufficiently high. This fact
gives us a lowerbound on the reheating temperature after the decay of
the Polonyi field.

The organization is as follows. In the next section, we show how the
baryon asymmetry of the universe can be explained if we assume the
Affleck-Dine mechanism for baryogenesis. In section~\ref{sec:lsp},
we calculate the mass density of LSP due to the decay of the Polonyi
field, and constrain the reheating temperature in the framework of the
minimal SUSY SU(5) model. Section~\ref{sec:discuss} is devoted to
discussion.

\section{Polonyi problem and the Affleck-Dine mechanism}
\label{sec:ad}

The Affleck-Dine mechanism~\cite{NPB249-361} for baryogenesis is based
on the fact that there are some combinations of squark $\tilde{q}$ and
slepton $\tilde{l}$ fields for which the scalar potential vanishes
identically when SUSY is unbroken. After SUSY breaking, these
flat-direction fields acquire masses $m_\chi$ of order $100\GEV$.  One
of these flat directions $\chi$ is assumed to have a large initial
value $\chi_0$ which is assumed to be about the grand unified theory
(GUT) scale $M_{GUT}\sim 10^{16}\GEV$ or the gravitational scale. It
has been shown~\cite{NPB249-361} that the decay of the coherent
oscillation mode of such a field $\chi$ can generate a large
baryon-to-entropy ratio $\sim O(1)$ under the presence of tiny
baryon-number violating operators such as
$(m_{S}/M_{GUT})\tilde{q}\tilde{q}\tilde{q}\tilde{l}$ (with $m_{S}$
being the scale of the SUSY breaking parameter in the observable
sector, which is assumed to be $m_{S}\sim O(100\GEV)$).

We now compute how large baryon asymmetry can be obtained if we
combine the Affleck-Dine mechanism with the Polonyi problem. For this
purpose, it is convenient to use the fact that $n_B/\rho_\phi$ is
independent of time since the baryon number is approximately conserved
in the regime we consider. (Here, $n_B$ is the baryon number density
and $\rho_\phi$ the mass density of the Polonyi field.) Then,
\begin{eqnarray}
	\frac{m_\chi n_B}{\rho_\phi} = {\rm const}.
\label{constant}
\end{eqnarray}
We evaluate this when the Affleck-Dine field $\chi$ starts its
oscillation. At this time,
\begin{eqnarray}
	\frac{m_\chi n_B}{\rho_\phi} \simeq \frac{m_\chi n_B}{\rho_\chi}
	\frac{\rho_\chi}{\rho_\phi}
	\simeq
	\eta_{B0} \frac{\rho_\chi}{\rho_\phi}
	\simeq
	\eta_{B0} \left( \frac{\chi_0}{\sqrt{3}M} \right)^2.
\label{at_oscillation}
\end{eqnarray}
where $\rho_\chi$ is the mass density of the Affleck-Dine field and
$\eta_{B0}\equiv (n_B/n_\chi )_{H\simeq m_\chi}$ with $n_\chi$ being
the number density of $\chi$. In deriving Eq.(\ref{at_oscillation}),
we have used $H\simeq\sqrt{\rho_\phi}/\sqrt{3}M$, and
$\rho_\chi=m_\chi^2\chi_0^2$. On the other hand, we evaluate the same
quantity given in Eq.~(\ref{constant}) at the decay time of the
Polonyi field $\phi$
\begin{eqnarray}
	\frac{m_\chi n_B}{\rho_\phi} \simeq \frac{4}{3}
	\frac{m_\chi n_B(T_R)}{s(T_R) T_R}.
\label{at_decay}
\end{eqnarray}
Equating Eq.~(\ref{at_oscillation}) and Eq.~(\ref{at_decay}), we get
\begin{eqnarray}
	\frac{n_B}{s} &\simeq&
        \frac{1}{4} \eta_{B0} \frac{T_R}{m_\chi}
	\lsp\frac{\chi_0}{M}\rsp^2
        \sim 10^{-5} \eta_{B0}
	\lsp\frac{T_R}{1\MEV}\rsp
	\lsp\frac{100\GEV}{m_\chi}\rsp
	\lsp\frac{\chi_0}{M}\rsp^2.
\label{eta_now}
\end{eqnarray}
With Eq.(\ref{eta_now}), one may explain the observed value $n_B/s
\sim (10^{-10} - 10^{-11})$ taking $\chi_0 \sim M_{GUT}$, $T_R\sim
1{\rm MeV}$, and $\eta_{B0}\sim O(1)$.\footnote
{It has been pointed out that the Affleck-Dine mechanism for
baryogenesis may result in too large baryon number fluctuation in the
case of chaotic inflation~\cite{APP2-291}. However, such a difficulty
can be solved if we adopt a larger value of the initial amplitude of
the Affleck-Dine field; $\chi_0\sim M$. In that case, we have to
choose $\eta_{B0}\sim 10^{-5}$.}

In our case, the dilution factor $D$ is given by
\begin{eqnarray}
	D \sim \frac{T_R}{m_\chi}
	\lsp\frac{\chi_0}{M}\rsp^2
	\sim
	10^{-5}
	\lsp\frac{T_R}{1\MEV}\rsp
	\lsp\frac{100\GEV}{m_\chi}\rsp
	\lsp\frac{\chi_0}{M}\rsp^2,
\label{dilution}
\end{eqnarray}
which is much larger than that derived in the previous
work~\cite{PLB174-176}. For example, the dilution factor given in
Ref.~\citen{PLB174-176} is $O(10^{-14})$ for the case $T_R\sim 1{\rm
MeV}$, which is about $10^{-9}$ times smaller than our result with
$m_\chi \sim 100{\rm GeV}$ and $\chi_0 \sim M$. This discrepancy
originates to the fact that the amplitude of the Polonyi field has
already decreased by a large amount at the decay time of the
Affleck-Dine field. In Ref.~\citen{PLB174-176}, this effect is not
taken into account, and hence the dilution factor given in
Ref.~\citen{PLB174-176} is underestimated.

\section{Mass density of LSP}
\label{sec:lsp}

Let us now turn to discuss a new cosmological difficulty in the
present solution to the Polonyi problem.  The decay of the Polonyi
field produces a large number of superparticles, which promptly decay
into LSPs. The number density of LSP produced by the decay, $n_{{\rm
LSP},i}$, is of the same order of that of the Polonyi field
$n_\phi\equiv\rho_\phi /m_\phi$. Just after the decay of the Polonyi
field, the yield variable for LSP, $Y_{\rm LSP}$, which is defined by
the ratio of the number density of LSP to the entropy density $s$, is
given by
\begin{eqnarray}
    m_{\rm LSP} Y_{\rm LSP} & \simeq & \frac{\rho_\phi}{s}
    \simeq \frac{m_{\rm LSP}\rho_{{\rm LSP},i}}{m_\phi s}
    \sim \frac{m_{\rm LSP}T_R}{m_\phi}
    \nonumber \\
    & \sim & 10^{-5}\GEV \left(\frac{m_{\rm LSP}}{100\GEV}\right)
    \left(\frac{T_R}{1\MEV}\right)\left(\frac{10\TEV}{m_\phi}\right),
    \label{mY_LSP}
\end{eqnarray}
where $\rho_{{\rm LSP},i}$ is the mass density of LSP just after the
decay of the Polonyi field. If LSP is stable and the pair annihilation
of LSP is not effective, $Y_{\rm LSP}$ is conserved until today.  On
the other hand, the ratio of the critical density $\rho_c$ to the
present entropy density $s_{0}$ is given by
\begin{equation}
    \frac{\rho_c}{s_{0}} \simeq 3.6 \times 10^{-9}h^2~\GEV,
    \label{critical}
\end{equation}
where $h$ is the Hubble constant in units of 100km/sec/Mpc.  Comparing
Eq.(\ref{mY_LSP}) with Eq.(\ref{critical}), we see that LSP overcloses
the universe in the wide parameter region for $m_{\rm LSP}, m_{\phi}$
and $T_R$ which we are concerned with.

If the pair annihilation of LSP takes place effectively, its abundance
is reduced to
\begin{equation}
    \frac{n_{\rm LSP}}{s} \simeq
    \left. \frac{H}{s\langle\sigma_{\rm ann}v_{\rm rel}\rangle}
      \rabs_{T=T_R},
      \label{abundance_LSP}
\end{equation}
where $\sigma_{\rm ann}$ is the annihilation cross section, $v_{\rm
rel}$ is the relative velocity, and $\langle\cdots\rangle$ represents
the average over the phase space distribution of LSP. Comparing
Eq.(\ref{critical}) with Eq.(\ref{abundance_LSP}), we obtain a
lowerbound on the annihilation cross section,
\begin{equation}
    \langle\sigma_{\rm ann}v_{\rm rel}\rangle \gsim
    3\times 10^{-8}h^{-2}\GEV^{-2}
    \left( \frac{m_{\rm LSP}}{100\GEV}\right)
    \left( \frac{100\MEV}{T_R}\right),
    \label{sv_limit}
\end{equation}
in order that the mass density of LSP does not overclose the universe.

As we can see, constraint (\ref{sv_limit}) becomes severer as the
reheating temperature $T_R$ decreases, and hence we obtain a
lowerbound on $T_R$. Here, we derive the constraint on $T_R$ in the
framework of minimal SUSY SU(5) model~\cite{NPB193-150,ZPC11-153},
which is shown in Appendix~\ref{ap:su5}.  We first solve RGEs based on
the minimal SU(5) model with the no-scale boundary conditions, and
determine the mass spectrum and mixing matrices of the superparticles.
Notice that we only investigate the parameter space which is not
excluded by the experimental or theoretical constraints.  The crucial
constraints are as follows;
\begin{itemize}
\item Higgs bosons $H_f$ and $\bar{H}_f$ have correct vacuum
expectation values; $\langle
H_f\rangle^2+\langle\bar{H}_f\rangle^2\simeq (174\GEV)^2$ and
$\tan\beta=\langle{H_f}\rangle/\langle{\bar{H}_f}\rangle$.
\item Perturbative picture is valid below the gravitational scale.
\item LSP is neutral.
\item Sfermions (especially, charged sleptons) have masses larger than
the experimental lower limits~\cite{PDG}.
\item The branching ratio for $Z$-boson decaying into neutralinos is
not too large~\cite{PLB350-109}.
\end{itemize}
One remarkable thing is that {\it LSP almost consists of bino which is
the superpartner of the gauge field for $U(1)_Y$} if we require that
LSP is neutral. Therefore, in our model, the LSP mass $m_{\rm LSP}$ is
essentially equivalent to the bino mass.  Then, we calculate the
annihilation cross section and determine the lowerbound on the
reheating temperature from the following equation;
\begin{equation}
    \left. \frac{H}{s\langle\sigma_{\rm ann}v_{\rm rel}\rangle}
      \rabs_{T=T_R} \leq
      \frac{\rho_c}{s_0} \simeq 3.6h^2 \times 10^{-9}\GEV.
\end{equation}

Since LSP is most dominated by bino, it annihilates into fermion
pairs. The annihilation cross section is given by~\cite{PLB230-78}
\begin{equation}
    \langle\sigma_{\rm ann}v_{\rm rel}\rangle
    = a + b\langle v^2\rangle,
    \label{sigma*v}
\end{equation}
where $\langle v^2\rangle$ is the average velocity of LSP,
and
\begin{eqnarray}
    a  & \simeq &
    \frac{32\pi\alpha_1^2}{27}
    \frac{m_t^2}{(m_{\tilde{t}_R}^2 + m_{\rm LSP}^2 - m_t^2)^2}
    \lsp 1 - \frac{m_t^2}{m_{\rm LSP}^2} \rsp^{1/2}
    \theta (m_{\rm LSP}-m_t),
    \label{s-wave} \\
    b
    &\simeq& \frac{8\pi\alpha_1^2}{3} \sum_{m_f\leq m_{LSP}} Y_f^4 \lmp
    \frac{m_{\rm LSP}^2}{(m_{\rm LSP}^2+m_{\tilde{f}}^2)^2}
    - \frac{2m_{\rm LSP}^4}{(m_{\rm LSP}^2+m_{\tilde{f}}^2)^3}
    + \frac{2m_{\rm LSP}^6}{(m_{\rm LSP}^2+m_{\tilde{f}}^2)^4} \rmp.
    \label{p-wave}
\end{eqnarray}
Here, $\alpha_1^2\equiv g_1^2/4\pi\simeq 0.01$ represents the fine
structure constant for U(1)$_{\rm Y}$, $m_t$ the top-quark mass, $Y_f$
the hypercharge of the fermion $f$, and $m_{\tilde{f}}$ the mass of
the sfermion $\tilde{f}$.  Notice that $a$- and $b$-terms correspond
to $s$- and $p$-wave contributions, respectively. Taking
$m_{\tilde{f}}\sim m_{\rm LSP}\sim 100\GEV$, the annihilation cross
section given in Eq.(\ref{sigma*v}) is at most $3\times
10^{-8}\GEV^{-2}$. Using this result in the inequality
(\ref{sv_limit}), we can see that the reheating temperature must be
higher than about 100MeV even if $\langle v^2\rangle\sim 1$. In fact,
LSP is in kinetic equilibrium in the thermal bath~\cite{lbl37715},
and hence its velocity is given by $O(T_R/m_{\rm LSP})$ which is much
smaller than 1.  Thus, we have severer constraint on $T_R$, as we will
see below.

In Fig.~1, we show the lowerbound on the reheating temperature in the
$\tan\beta$ vs. $m_{\rm LSP}$ plane. In the figures, large or small
$\tan\beta$'s are not allowed since the Yukawa coupling constant for
the top quark or bottom quark blows up below the gravitational scale
for such $\tan\beta$'s. Furthermore, there also exists a lowerbound on
the LSP mass. In the case where $\tan\beta\lsim 20$, charged sfermions
become lighter than the experimental limit if the LSP mass becomes
lighter than $\sim 50\GEV$.  On the other hand, for the large
$\tan\beta$ case, unless the bino mass is sufficiently large, the
lightest charged slepton becomes LSP. (Remember that the dominant
component of LSP is bino.)  Thus, the lowerbound on $m_{\rm LSP}$ is
obtained. As we can see, the reheating temperature should be larger
than about 100MeV, even for the case where $m_{\rm LSP}\sim 50\GEV$.
The constraint becomes more stringent as $m_{\rm LSP}$ increases,
since the masses of the superparticles which mediate the annihilation
of LSP becomes larger as the LSP mass increases. If we translate the
lowerbound on the reheating temperature into that of the Polonyi mass
$m_\phi$, we obtain $m_\phi\gsim 100\TEV$ (see
Eq.(\ref{rtemp})).

Finally, we comment on the accidental case where the annihilation
process hits the Higgs pole in the $s$-channel. If the LSP mass is
just half of the lightest Higgs boson mass, the LSP annihilation cross
section is enhanced since LSP has small but nonvanishing fraction of
higgsino component. If the parameters are well tuned, such a situation
can be realized and the lowerbound of $T_R$ decreases to $O(10\MEV)$.
However, we consider that such a scenario are very unnatural since a
precise adjustment of the parameters is required in order to hit the
Higgs pole. \footnote
{In the case where the annihilation process hits the pole of heavier
Higgs bosons, the cross section is not enhanced so much, since the
widths of the heavier Higgs bosons are quite large.}

\section{Discussion}
\label{sec:discuss}

Here, we proposed a solution to the Polonyi problem based on the
no-scale type supergravity and the Affleck-Dine mechanism for
baryogenesis. In our scenario, however, LSP may be overproduced due to
the decay of the Polonyi field. From this fact, we obtained the
lowerbound on the reheating temperature after the decay of the Polonyi
field, which is given by $O(100\MEV)$. As a result, the mass of the
Polonyi field have to be larger than $O(100\TEV)$, which may raise a
new fine-tuning problem~\cite{PLB173-303,PLB168-347}.

To cure this conflict in the case of $T_R \lsim 10$ MeV, let us
consider modifications of the minimal SUSY standard model (MSSM).  One
way is to extend the particle contents and provide a new, very light
LSP.  If the LSP is lighter than $O(10$ MeV), we can see from
Eq.~(\ref{mY_LSP}) that the relic abundance does not exceed the
critical density without invoking the annihilation.  This is most
easily realized in the minimal extension of the MSSM, where the
superpartner of a singlet Higgs is contained in the neutralino sector.
Another extension which has a light LSP is to incorporate the
Peccei-Quinn symmetry.  Then the superpartner of the axion, the axino,
can be the LSP~\cite{NPB358-447}.  Indeed, it was shown in
Ref.~\citen{PLB276-103} that the axino becomes massless at the
tree-level in the no-scale supergravity.  Radiative corrections may
give a small, model-dependent axino mass.\footnote
{A light axino can also be realized if one chooses a special form of
superpotential~\cite{PLB287-123}.}
In the case of the axino mass $\sim 10\MEV$, the axino becomes a cold
dark matter of the universe.

$R$-parity breaking is the other possibility to make our scenario
cosmologically viable. In this case, the LSP is no longer stable, but
decays to ordinary particles.  If the lifetime $\tau_{LSP}$ of the LSP
is shorter than 1~sec,\footnote
{Such a small $R$-parity violation ($\tau_{LSP} \sim 1{\rm sec}$) is
consistent with other phenomenological constraints~\cite{PLB256-457}.}
its decay does not upset the standard big-bang nucleosynthesis.

\section*{Acknowledgement}

The author would like to thank M.~Kawasaki, M.~Yamaguchi and
T.~Yanagida for useful discussion, and J.~Yokoyama for a comment on
the Affleck-Dine mechanism for baryogenesis.

\appendix

\section{The model}
\label{ap:su5}

In this appendix, we describe the model we use, {\it i.e.} the minimal
SUSY SU(5) model~\cite{NPB193-150,ZPC11-153} with no-scale type
boundary conditions. This model has three types of Higgs field;
$H({\bf 5})$ and $\bar{H}({\bf 5^*})$ which contain flavor Higgses
$H_f$ and $\bar{H}_f$, and $\Sigma ({\bf 24})$ whose condensation
breaks the SU(5) group into the gauge group of MSSM, $\rm
SU(3)_C\times SU(2)_L\times U(1)_Y$. For the Higgs sector, the
superpotential is given by
\begin{equation}
    W = \frac{1}{3} \lambda {\rm tr} \Sigma^3
    + \frac{1}{2} M_\Sigma {\rm tr} \Sigma^2
    + \kappa \bar{H} \Sigma H
    + M_H \bar{H} H,
    \label{W_su5}
\end{equation}
where $\lambda$ and $\kappa$ are dimensionless constants, while
$M_\Sigma$ and $M_H$ are mass parameters which are of the order of the
grand unified theory (GUT) scale $M_{\rm GUT} (\sim 10^{16}\GEV)$.
Furthermore, the model also has the soft SUSY breaking terms;
\begin{equation}
    {\cal L}_{\rm soft} =
      - \frac{1}{3} \lambda A_\Sigma {\rm tr} \Sigma^3
      - \frac{1}{2} M_\Sigma B_\Sigma {\rm tr} \Sigma^2
      - \kappa A_H \bar{H} \Sigma H
      - M_H B_H \bar{H} H +h.c.,
    \label{L_soft_su5}
\end{equation}
where $A_\Sigma$, $B_\Sigma$, $A_H$ and $B_H$ are SUSY breaking
parameters.  Minimizing the Higgs potential, we find the following
stationary point;
\begin{equation}
    \vev{\Sigma} =
    \frac{1}{\lambda} \lmp
    M_\Sigma + 2 \lsp A_\Sigma - B_\Sigma \rsp
    + O\lsp \frac{A_\Sigma}{M_\Sigma}, \frac{B_\Sigma}{M_\Sigma} \rsp
    \rmp \times {\rm diag}(2,2,2,-3,-3),
    \label{vac_su5}
\end{equation}
where the SU(5) is broken down to $\rm SU(3)_C\times SU(2)_L\times
U(1)_Y$. Regarding this stationary point as the vacuum, we obtain MSSM
as the effective theory below the GUT scale $M_{\rm GUT}$. Here, the
masslessness of the flavor Higgses $H_f$ and $\bar{H}_f$ is achieved
by a fine tuning among several parameters; $M_H - 3 \kappa
M_\Sigma/\lambda \simeq \mu_H$, where $\mu_H$ is the SUSY-invariant
Higgs mass in MSSM.

In the present model, the parameters in MSSM at the electroweak scale
is obtained by solving renormalization group equations (RGEs).  The
boundary conditions on the parameters in the minimal SUSY SU(5) model
are given at the gravitational scale $M$.  Since we assume the
no-scale type supergravity models, all the SUSY breaking parameters
except for the gaugino mass vanish at the gravitational scale. From
the gravitational scale to the GUT scale, the parameters follow the
renormalization group flow derived from RGEs in the minimal SUSY SU(5)
model. Then we determine the parameters in MSSM at the GUT scale
through an appropriate matching condition between the parameters in
the SUSY SU(5) model and those in MSSM.  Finally, we use RGEs in MSSM
from the GUT scale to the electroweak scale in order to obtain the low
energy parameters.

As for the matching condition, we have a comment. In the stationary
point (\ref{vac_su5}), the mixing soft mass term of the two flavor
Higgs bosons, $m_{12}^2\bar{H}_fH_f$, is generated at the tree level,
where $m_{12}^2$ is given by
\begin{equation}
    m_{12}^2 (M_{\rm GUT}) \simeq
      \llp \frac{6\kappa}{\lambda}
      (A_\Sigma - B_\Sigma)(A_H - B_\Sigma)
      -\mu_H B_H
      \rlp_{\mu =M_{\rm GUT}}.
      \label{m12^2}
\end{equation}
Since the mixing mass term depends on unknown parameters, $\lambda$
and $\kappa$ in Eq.(\ref{W_su5}), we regard $m_{12}^2$ as a free
parameter taking account of the uncertainty of $\lambda$ and $\kappa$
in our analysis.  Then, the low energy parameters are essentially
determined by the gauge and Yukawa coupling constants and the
following three parameters; the supersymmetric Higgs mass $\mu_H$, the
mixing mass of the two flavor Higgs bosons $m_{12}^2$, and the unified
gaugino mass.\footnote
{In fact, parameters in MSSM slightly depend on the parameters in the
SUSY GUT such as $\lambda$, $\kappa$ and so on. In our numerical
calculation, we ignore the effects of these parameters on the
renormalization group flow.}
However, it is more convenient to express these parameters by other
physical ones. In fact, one combination of them is constrained so that
the flavor Higgs bosons have correct vacuum expectation values;
$\langle H_f\rangle^2+\langle\bar{H}_f\rangle^2\simeq (174\GEV)^2$. As
the other two physical parameters, we use the mass of LSP, $m_{\rm
LSP}$, and the vacuum angle
$\tan\beta\equiv\langle{H_f}\rangle/\langle{\bar{H}_f}\rangle$. Thus,
once we fix $m_{\rm LSP}$ and $\tan\beta$, we can determine all the
parameters in MSSM.\footnote
{Yukawa coupling constants are determined so that the fermions have
correct masses. The gauge coupling constants are also fixed so that
their correct values at the electroweak scale are reproduced.}

\newpage

\begin{figure}[p]
\centerline{\psfig{figure=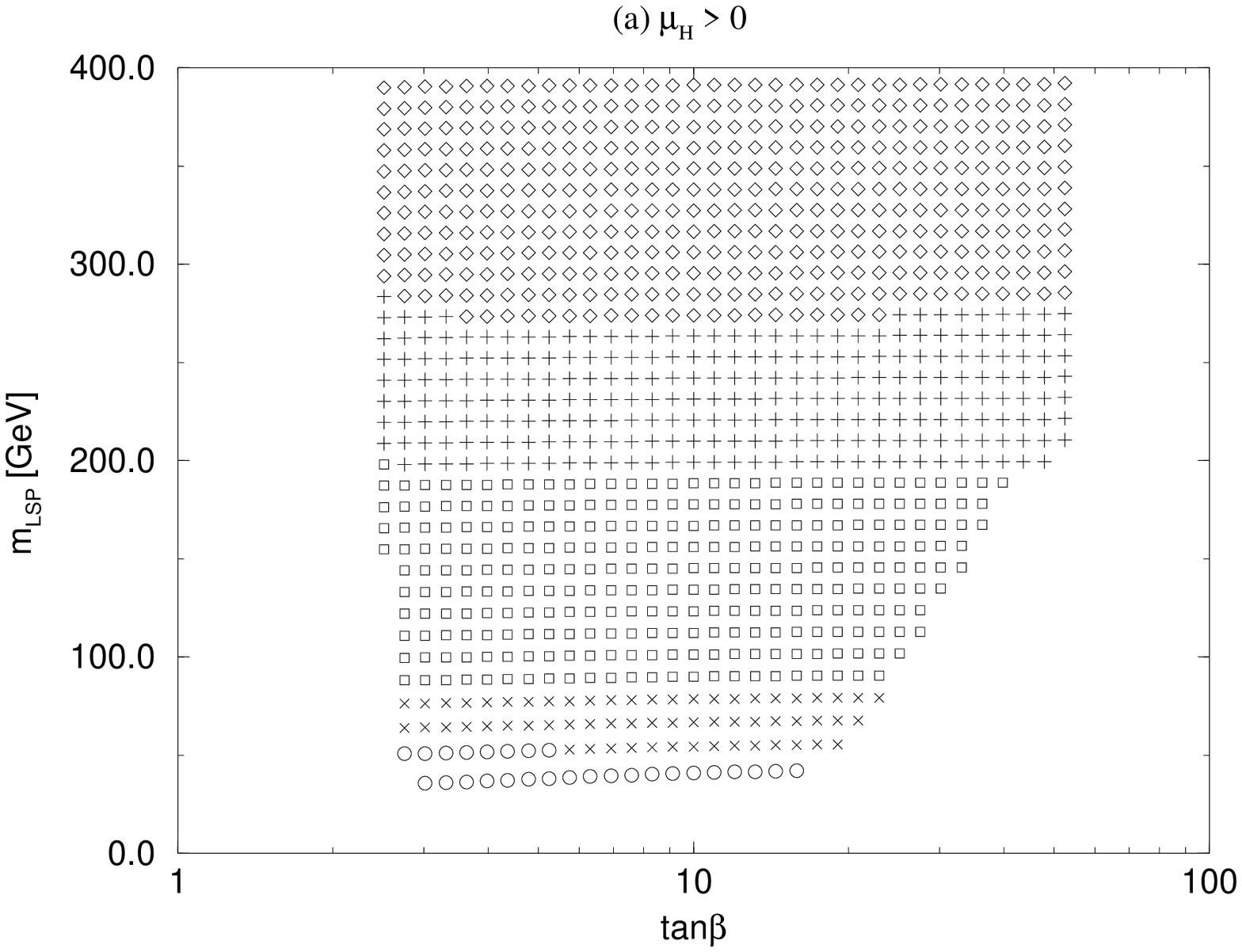,%
bbllx=156bp,bblly=292bp,bburx=406bp,bbury=563bp,width=6cm,angle=-90}}
\vskip 1.6in
\centerline{\psfig{figure=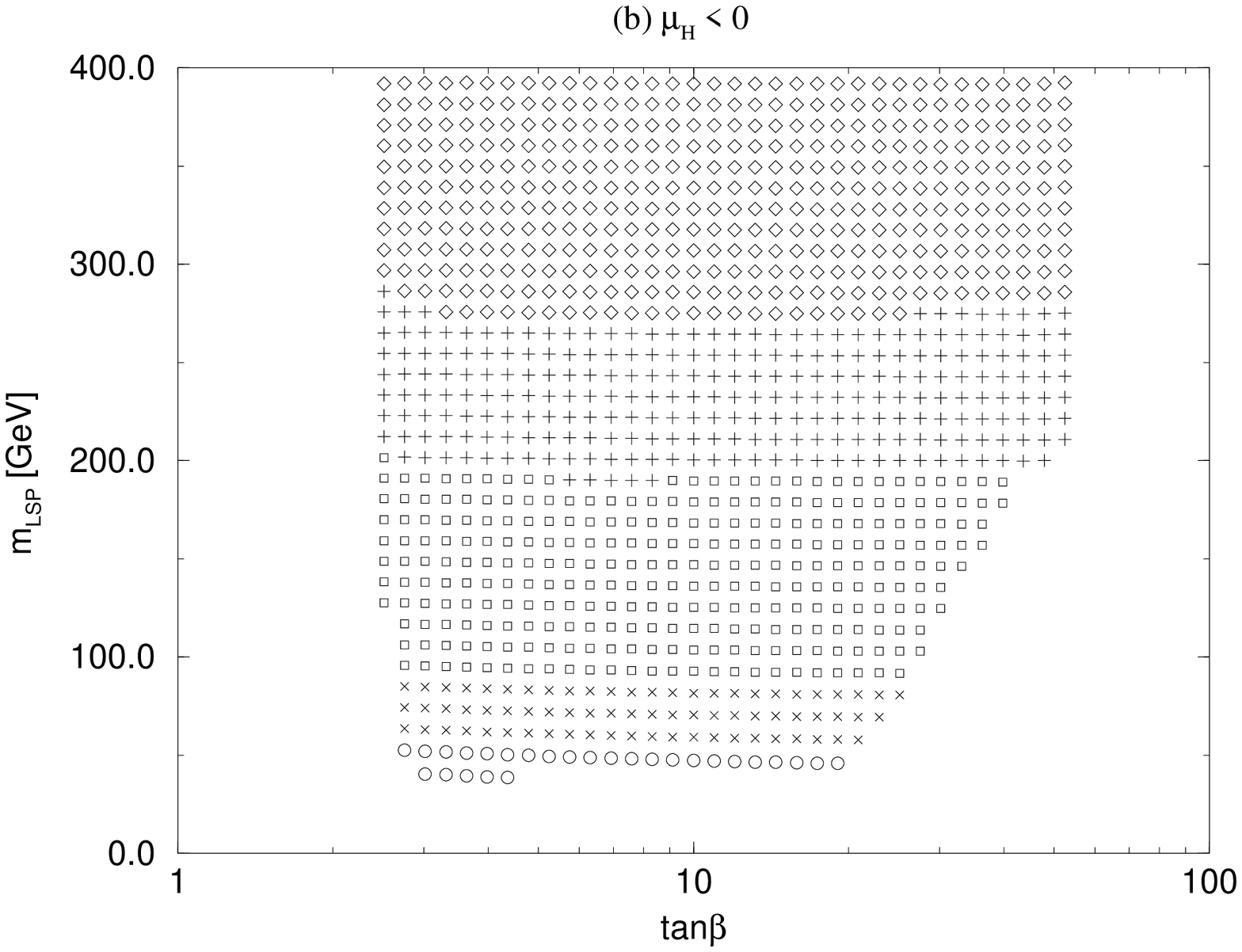,%
bbllx=156bp,bblly=292bp,bburx=406bp,bbury=563bp,width=6cm,angle=-90}}
\vskip 1.2in
\caption{Lowerbound on $T_R$ is shown in $\tan\beta$ vs. $m_{\rm LSP}$
plane. The meaning of each mark is as follows; $\circ : 100\MEV\leq
T_R\leq 500\MEV$, $\times : 500\MEV\leq T_R\leq 1\GEV$, $\Box :
1\GEV\leq T_R\leq 5\GEV$, $+ : 5\GEV\leq T_R\leq 10\GEV$, $\Diamond :
10\GEV\leq T_R\leq 50\GEV$. The sign of the SUSY-invariant Higgs mass
$\mu_H$ is taken to be (a) $\mu_H >0$, and (b) $\mu_H <0$.}
\end{figure}

\end{document}